\documentclass[useAMS,usenatbib]{mn2e}
\usepackage{graphicx}
\def\eq{\begin{equation}}         % Forma curta do ambiente \begin{equation}
\def\eeq{\end{equation}}      % Idem, \end{equation}
\title[WX~Cen -- a possible type Ia Supernova progenitor]{WX~Cen ($\equiv$ WR~48c) -- a possible
 type Ia Supernova progenitor\thanks{
Based on observations made at Laborat\'orio Nacional de Astrof\'{\i}sica/CNPq, Brazil, and at the 1.5m 
	ESO telescope at La Silla, Chile.}}
\author[A. S. Oliveira and J. E. Steiner]{A. S. Oliveira \thanks{E-mail:
alex@astro.iag.usp.br (ASO); steiner@astro.iag.usp.br (JES)},
and J. E. Steiner \\
Instituto de Astronomia, Geof\'{\i}sica e Ci\^encias Atmosf\'ericas - IAG, Universidade de S\~ao Paulo, CP 9638, 01065-970,
               S\~ao Paulo, Brazil\\}
\begin{document}

\date{Accepted ????. Received ????; in original form ????}

\pagerange{\pageref{firstpage}--\pageref{lastpage}} \pubyear{2003}

\maketitle

\label{firstpage}

\begin{abstract}

We confirm the orbital period of WX~Cen $\equiv$ WR~48c determined by \citet{dia95} -- DS95 -- 
and refined its value to $P_{orb} = 0.416~961~5 (\pm 22)$ d. The light curve of 
this object has a peak to peak 
variation of approximately 0.32 magnitudes. It is non-sinusoidal in the sense that it has a V-shaped narrow minimum,
similar to the ones seen in V~Sge, V617~Sgr and in 
Compact Binary Supersoft Sources -- CBSS. 

Most of the emission lines in the optical spectrum are due to Balmer, He II, C IV, N V, 
O V and O VI. An analysis of the He II Pickering series decrement shows that the system 
has significant amount of hydrogen. The emission lines of He II 4686{\AA} became weaker 
between the 1991 and 2000/2002 observations, indicating distinct levels of activity. The spectra of WX~Cen show variable absorption 
features in the Balmer lines with 
$V = -2900$ km~s$^{-1}$ and in emission with $V = \pm 3500$ km~s$^{-1}$. These highly variable events remind the 
satellites in emission of CBSS. 

We estimate the color excess as $E(B-V)=0.63$  on the basis of the observed diffuse interstellar band at 
5780{\AA}. Given the distance-color 
excess relation in the direction of WX Cen, this implies a distance of $2.8  \pm 0.3$ kpc. Interstellar absorption of the 
Na I D lines show components at $-4.1$ km~s$^{-1}$, which corresponds to the velocity of the Coalsack, and three other 
components a $-23.9$, $-32.0$ and $-39.0$ km~s$^{-1}$. These components are also seen with similar strengths in field stars 
that have distances between 1.8 and 2.7 kpc. The intrinsic color of WX Cen is $(B-V)_0=-0.2$ and the absolute 
magnitude, $M_V = -0.5$.

Extended red wings in the strong emission lines are seen. A possible explanation is that  the system has a spill-over 
stream similar to what is seen in V617~Sgr. We predict that when observed in opposite phase, blue wings would 
be observed. A puzzling feature that remains to be explained is the highly variable red wing ($V \sim 700$ km~s$^{-1}$) 
of the O VI emission lines as well as of the red wings of the H and He lines.

The velocity of the satellite-like feature is consistent with the idea that the central star is a white dwarf with a mass of 
$M \sim 0.9 M_{\sun}$. With the high accretion rate under consideration, the star may become a  SN~Ia in a time-scale of 
$5 \times 10^{6}$ years.

\end{abstract}

\begin{keywords}
stars: individual: WX Cen -- stars: binaries: close -- stars: emission-line.
\end{keywords}

\section{Introduction}

WX~Cen was initially identified by \citet{egg} as a possible optical counterpart of the hard X-ray transient 
source Cen XR-2, although this identification was latter discarded.
Because of its spectral characteristics, the object was then classified as a Wolf-Rayet star of type 
WN 7 \citep{huc81}, while \citet{vogt} classified it as a nova-like 
variable. \citet{dia95} -- DS95 -- showed that it is a binary system with an orbital period of 10.0 hr. The 
photometric orbital variation, determined by these authors, based on their spectroscopic observations, is approximately 
sinusoidal with an amplitude of $\sim 0.3$ mag. In spite of the period being relatively long (in the context of Cataclysmic 
Variables) it was not possible to identify any spectral signature of the secondary star. 
A distance of 1400 pc was determined from the Na I D line equivalent width, and an $E(B-V) = 0.4 \pm 0.1$ was 
suggested from the average of three distinct determinations (4430{\AA} DIB and Na I D equivalent widths and He II 4686, 10124 {\AA}
lines ratio). Doppler tomography of WX~Cen 
produced by DS95 suggests that the gas of the secondary has normal chemical abundance, while the wind of the primary 
star is probably over-abundant in helium (He$^{++}$/H = 0.56). DS95 also showed that, if the primary component of WX~Cen is a white 
dwarf, then the secondary should have a mass bellow $0.35 M_{\sun}$ and, therefore, be evolved. If, however, one assumes
 the hypothesis that the secondary is a main sequence star, it would have $M_2 = 1.16 M_{\sun}$, and the primary, a mass of 
 $M_1 > 3.5 M_{\sun} \pm 0.5 M_{\sun}$. 

\citet{steiner98} included WX~Cen in a group of 4  galactic binaries defined as the V~Sge stars. The other three objects are
V~Sge \citep{her,dia99}, V617~Sgr \citep{steiner99,cie} and DI~Cru \citep{vee02c}. 
They are characterized by the presence of strong emission lines of O VI and N V. Besides, He II is 
at least two times more intense than H$\beta$. The V~Sge stars are similar to the Compact Binary Supersoft Sources 
(CBSS), seen in the Magellanic Clouds, but not that frequent in the Galaxy. CBSSs are interpreted as suffering 
hydrostatic hydrogen nuclear burning on the surface of a white dwarf. This burning is due to the high mass-transfer 
rate, a consequence of the fact that the systems have inverted mass ratios (see \citet{kah} for a revision and references). 
\citet{patt} also considered the hypothesis of Galactic CBSS nature for the stars V Sge, T Pyx and WX Cen, established on 
their extremely blue colors, high luminosities, orbital light curves and highly excited emission line spectra.

In an observational program to search for galactic WR stars, \citet{shara} rediscovered WX~Cen as a new WN3, 
when this star received the WR~48c designation. In our search for V~Sge type stars, we selected WR~48c as a candidate, on the basis of  
spectroscopic criteria, for a detailed observational study. We, then, realized that WR~48c was the already known star WX~Cen.

\begin{figure}
\vspace*{10pt}
\centerline{\includegraphics[width=84mm]{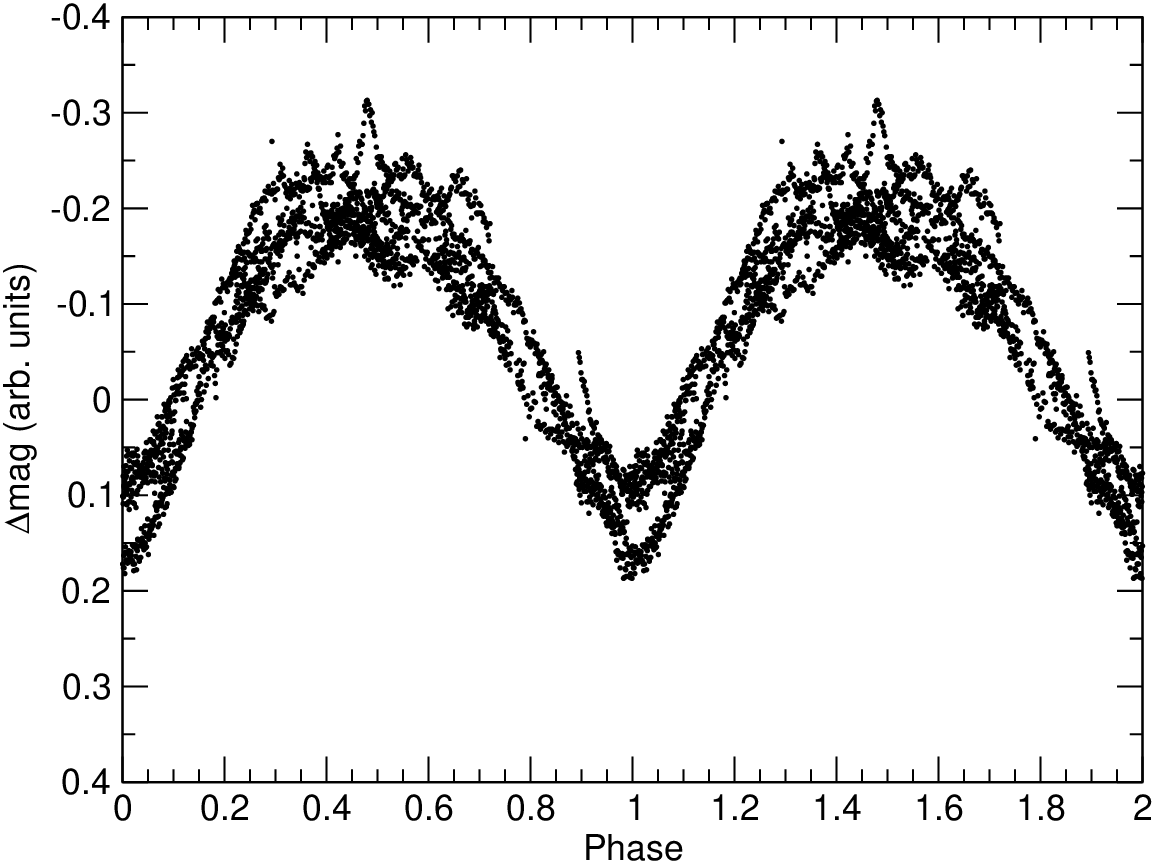}}
\caption{Average \textit{V} light curve of the data obtained in 2000, folded with the orbital period and epoch
	from the photometric ephemeris. \label{lca}}
\end{figure}

\begin{figure}
\vspace*{50pt}
\centerline{\includegraphics[width=84mm]{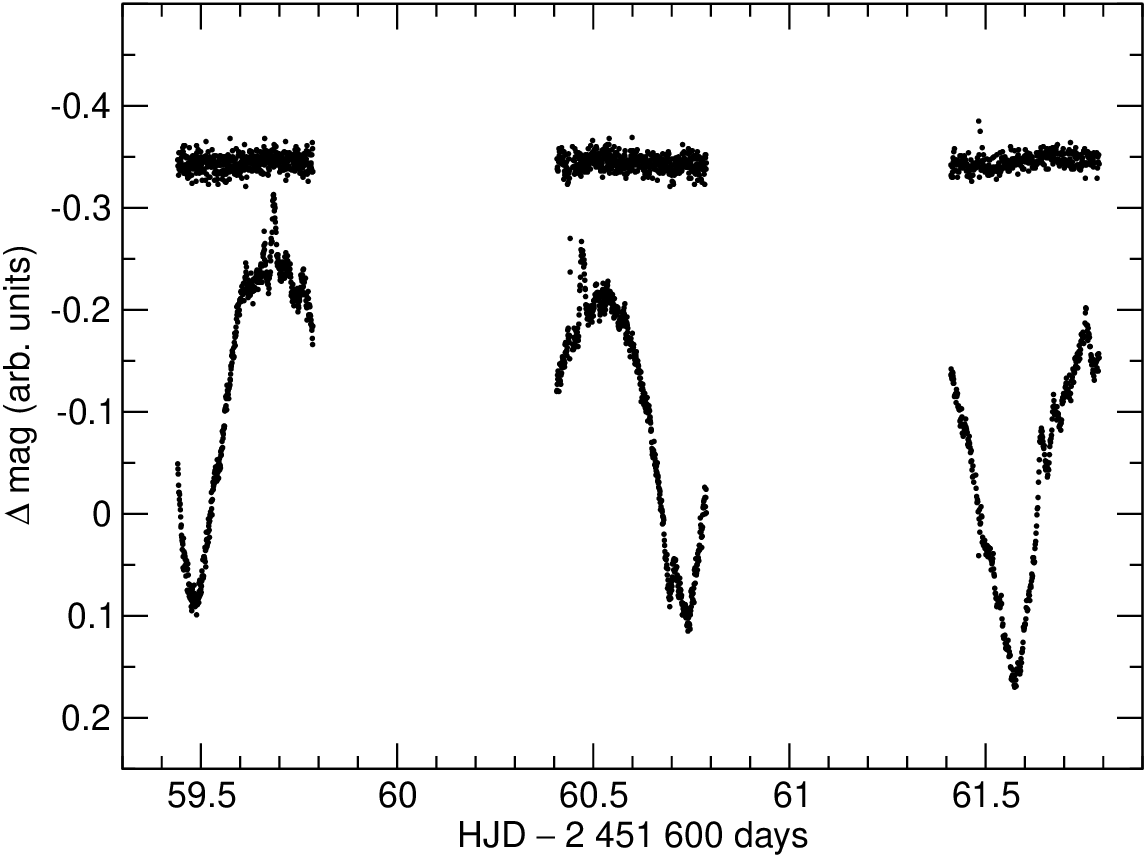}}
\caption{The light curve of WX~Cen for the nights 2000 April 24, 25 and 26. The upper light curves are
	of the comparison star. An increase of 0.08 mag in the mean magnitude is seen in consecutive nights. \label{lcb}}
\end{figure}

\section{Observations}\label{obs}

We observed WX~Cen photometrically for 9 nights on 2000 April and 2003 May and July with the Boller \& Chivens 60~cm and Zeiss 60~cm
telescopes at the Pico dos Dias Observatory -- MCT/LNA -- located in Braz\'opolis, Southeast Brazil (see Table~\ref{jophoto}).
A thin back-illuminated EEV CCD 002-06 chip and a Wright Instruments thermo-electrically cooled camera were employed to obtain the images.
 The timing was 
provided by a Global Positioning System (GPS) receiver. The frames were obtained through the Johnson $V$ 
filter, with exposure times of 40, 50 and 80 seconds. We also observed this star in white light with 2 seconds exposure time, for which 
we used the same CCD, operated in the frame transfer mode,
resulting in dead time between integrations of the order of milliseconds. Bias and flatfield exposures were obtained for
correction of the stellar images. The images, except the ones obtained in frame transfer mode, present
 an overscan region used for the CCD's zero
point correction. The data reduction was performed with the standard procedures, using IRAF~\footnote
{IRAF is distributed by the National Optical Astronomy Observatories,
which are operated by the Association of Universities for Research in Astronomy, Inc., under cooperative agreement 
with the National Science Foundation.}
routines. The star is located in a quite rich field; we, therefore, carried out differential PSF (Point Spread Function) 
photometry of WX~Cen and four comparison stars of the field using the LCURVE package, 
written and kindly provided by M.P. Diaz. It makes use of DAOPHOT routines to treat automatically long 
time series data. 

\begin{table}
 \centering
 %\begin{minipage}{140mm}
  \caption{Journal of photometric observations of WX~Cen.\label{jophoto}}
  \begin{tabular}{@{}lcccc@{}}
  \hline
   Date   & Telescope   &  Number  & Exp. time & Filter \\
  (UT)    & 	   &   of exps. & (s) &  \\
  \hline
2000 Apr 07	& Zeiss	& 461	& 40	& $V$ \\
2000 Apr 08	& Zeiss	& 386	& 40	& $V$\\
2000 Apr 09	& Zeiss	& 492	& 40	& $V$\\
2000 Apr 24	& B\&C	& 507	& 50	& $V$\\
2000 Apr 25	& B\&C	& 555	& 50	& $V$\\	
2000 Apr 26	& B\&C	& 372	& 80	& $V$\\
2003 May 20 	& Zeiss	& 327	& 50	& $V$\\
2003 May 21	& Zeiss	& 590	& 50	& $V$\\
2003 Jul 02	& Zeiss	& 5299	& 2	& clear\\
\hline
\end{tabular}
%\end{minipage}
\end{table}

We also observed the object with the Cassegrain spectrograph coupled to the 1.6 m Boller \& Chivens telescope at 
LNA on 2000 March. We used a dispersion grating with 900 lines per mm, covering the spectral range from 3700{\AA} to 4950{\AA} to
obtain spectra with exposure times of 10 minutes (see Table~\ref{jospec}). A thin back-illuminated SITe SI003AB 
1024x1024 CCD was used as detector. Bias and dome flatfield exposures were obtained for 
correction of the CCD read-out noise and sensitivity. The width of the slit was adjusted to the conditions of the seeing. 
We took exposures of calibration lamps after every third exposure of the star, in order to determine the wavelength calibration solution. 
The image reductions, spectra extractions and wavelength calibrations were executed with IRAF standard routines. 

\begin{figure}
\vspace*{15pt}
\centerline{\includegraphics[width=84mm]{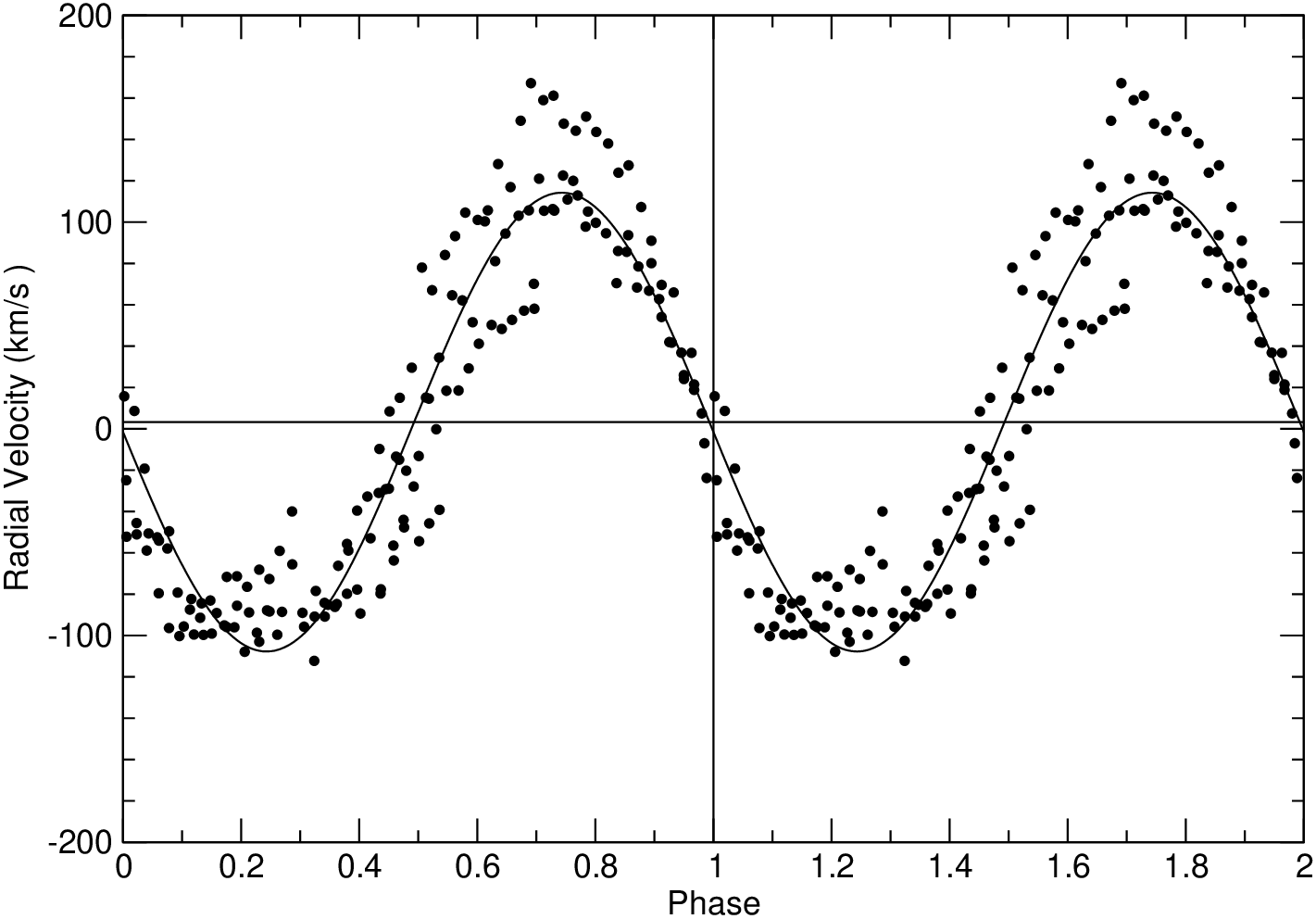}}
\caption{He II 4686{\AA} radial velocity curve, folded with the period and epoch given in the spectroscopic
	ephemeris. The sinusoidal curve fitted to the data has amplitude of 111 km~s$^{-1}$. \label{vr} }
\end{figure}

\begin{figure}
\vspace{50pt}
\centerline{\includegraphics[width=84mm]{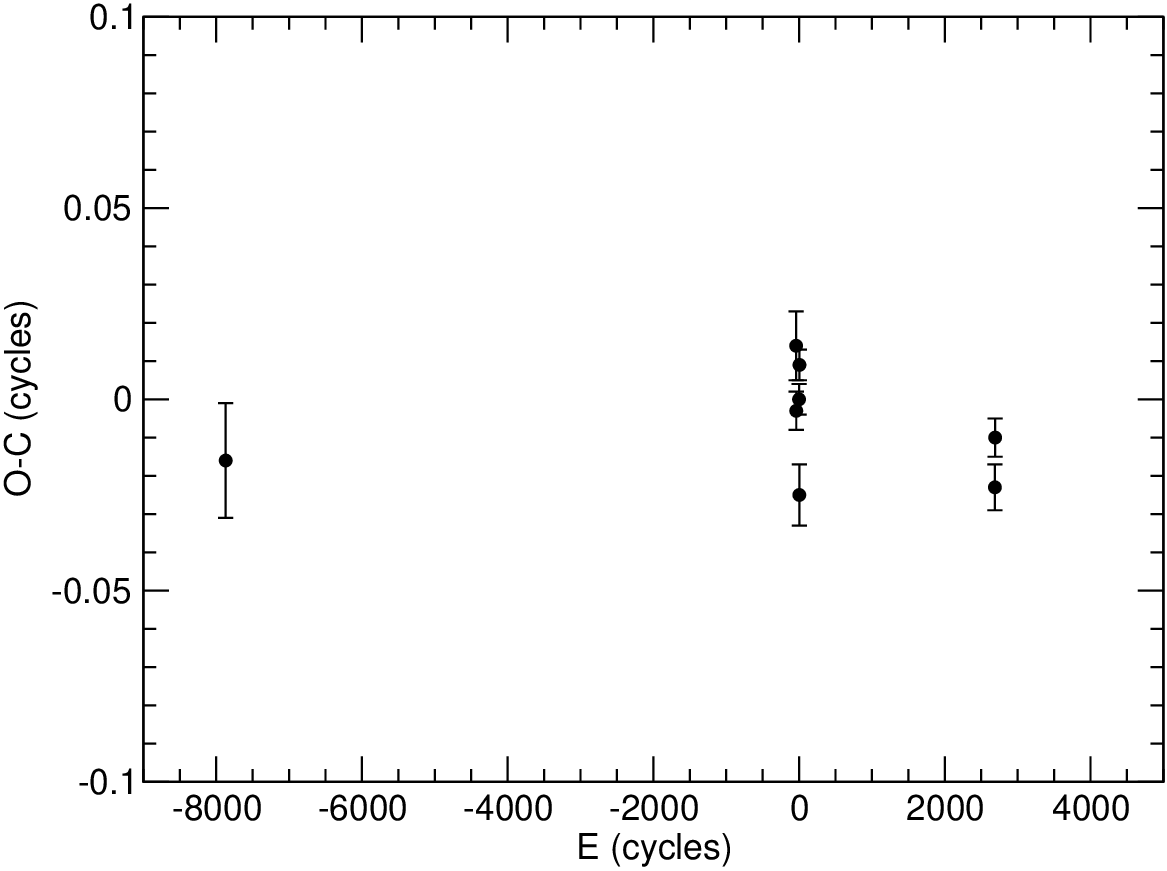}}
\caption{The O$-$C diagram corresponding to the photometric ephemeris. The first point (cycle $-7869$) is the timing of the minimum of the 
	continuum modulation presented by DS95. The other points are the timings of minima measurements 
	from the photometric data presented in this work (see Table~\ref{timing}). \label{o-c}}
\end{figure}

An additional spectroscopic observation was made with the FEROS (Fiber-fed Extended Bench Optical 	
Spectrograph - see \citet{kauf}) spectrograph at the 1.52 m telescope of ESO (European Southern Observatory) in 
La Silla, Chile, on 2002 January 24. The FEROS spectrograph uses a bench mounted Echelle grating with reception fibers 
in the Cassegrain focus. It supplies a resolution of $R=48~000$, corresponding to 2.2 pixels of 15 $\mu$$m$, and spectral 
coverage from 3600{\AA} to 9200{\AA}. A completely automatic on-line reduction system is available and was adopted by us. 
We obtained a single spectrum with integration time of 30 minutes.

\begin{table}
 \centering
% \begin{minipage}{140mm}
  \caption{Journal of spectroscopic observations of WX~Cen.\label{jospec}}
  \begin{tabular}{@{}llccc@{}}
  \hline
   Date     & Instrum.   &  Number  & Exp. time  & Resolution \\
   (UT)	&		 &of exps.	& (s) &({\AA})\\
  \hline
2000 Mar 12	& Cass.      & 6     & 600   & 2.8   \\
2000 Mar 24	& Cass.      & 53    & 600   & 2.8   \\
2000 Mar 25	& Cass.      & 54    & 600   & 2.8   \\
2000 Mar 26	& Cass.      & 54    & 600   & 2.8   \\
2002 Jan 24	& FEROS     & 1     & 1800  & 0.1   \\
\hline
\end{tabular}
%\end{minipage}
\end{table}

\section{Data analysis}

\subsection{The optical light curve}

The average light curve of WX~Cen (Fig.~\ref{lca}) displays a well defined 
orbital modulation with total amplitude of approximately 
0.32 magnitudes. It is asymmetric, and can not be adjusted by a simple sine wave. The maximum is wider and
the minimum, narrower than the expected from a sine wave. The rise is also faster than the fall to the minimum. This light 
curve has a minimum with a shape that resembles the ones of V~Sge and of V617~Sgr as well as the galactic CBSS QR And. Superposed on 
it, there is a fluctuation with time-scale of tens of minutes. In addition to this fast fluctuation, there is a variation with time-
scale of one day and with approximately 0.08 mag,
 as can be seen in the light curves of the nights of 2000 April 24, 25 and 26 (Fig.~\ref{lcb}). 
Single flare-like events with amplitude of 0.08 magnitudes and lasting for about 30 minutes are also seen (see Fig.~\ref{lcb}).

We searched the data for optical pulsations in the range of 10 to 200 cycles day$^{-1}$. No periodic signal is 
seen but power excess with $T \sim 85$ cycles day$^{-1}$ (17 min) is visible. This is to be compared to the 17--27 min 
oscillations, seen in V617 Sgr \citep{steiner99} and to the 60 min oscillations, seen in V Sge \citep{her}. 
In addition, a power law signal with 

\eq
Log~P = \alpha ~log f + C
\eeq

\noindent is seen, where $\alpha = -1.85 \pm 0.04$. This is the classical signature for low frequency flickering.

In order to estimate the inclination of the system, we will compare the light curve of WX Cen with that of QR And. 
From the observed amplitude of the light curve of QR And, which varies from $\Delta$m = 0.25 magnitudes \citep{meyer2}
to $\Delta$m = 0.5 magnitudes \citep{grath}, an inclination  of 
$i=55\degr$ was derived \citep{meyer2}. As the observed  light curve amplitude of  
WX Cen is within the  observed range for QR And, and as the overall light curve shape and photometric behavior 
are similar, we will adopt $i=55\degr$ for WX Cen as well.

\subsection{Refining the orbital period} \label {refi}

The orbital radial velocity curve obtained from the He II 4686{\AA} line is shown in Fig.~\ref{vr}. The spectroscopic 
ephemeris associated to this radial velocity curve is 

\eq
\footnotesize T_{(spec)} (HJD) = 2~451~659.544(\pm17) + 0.416~961~5(\pm22) \times E
\eeq

This ephemeris was calculated defining the zero phase as the crossing from positive to negative values of the radial velocities 
when compared to the systemic velocity, $\gamma = 3$ km~s$^{-1}$. We refined the spectroscopic period
from the value determined by DS95, as the number of orbital cycles elapsed from their zero phase timing to the 
present one is long enough, that is, 7870 cycles. 
	
The ephemeris of the light curve minima is 

\eq
\footnotesize T_{(min)} (HJD) = 2~451~659.485(\pm13) + 0.416~960~1(\pm28) \times E
\eeq

The value of the photometric period presented in this ephemeris was determined from the (O$-$C)
diagram shown in Fig.~\ref{o-c}, which was in turn constructed from the measurements presented 
in Table~\ref{timing}. The orbital period derived from the radial velocities and from the photometric
minima are the same, within 1$\sigma$ confidence level.	
The delay between the spectroscopic ephemeris and what is expected from the photometric one is 
about $\Delta\phi = 0.14$ and is similar to what is reported in DS95.	

\begin{table}
 \centering
% \begin{minipage}{140mm}
  \caption{Times of photometric minima.\label{timing}}
  \begin{tabular}{@{}lcc@{}}
  \hline
   Date     & Observed Minimum  &  E   \\
  (UT)	&(HJD)	&(cycles)	\\
  \hline
1991 Apr 31	& 2~448~378.410   ($\pm15$)	& -7869		\\
2000 Apr 07	& 2~451~642.821	($\pm9$)		& -40		\\ 	
2000 Apr 08	& 2~451~643.638	($\pm5$)		& -38		\\
2000 Apr 24	& 2~451~659.485	($\pm4$)		& 0		\\
2000 Apr 25	& 2~451~660.711	($\pm8$)		& 3		\\	
2000 Apr 26	& 2~451~661.579	($\pm4$)		& 5		\\
2003 May 20 	& 2~452~780.668	($\pm6$)		& 2689		\\
2003 May 21	& 2~452~781.515	($\pm5$)		& 2691		\\
\hline
\end{tabular}
%\end{minipage}
\end{table}

\subsection{The high-resolution spectrum}

\begin{figure}
\vspace*{50pt}
\centerline{\includegraphics[width=84mm]{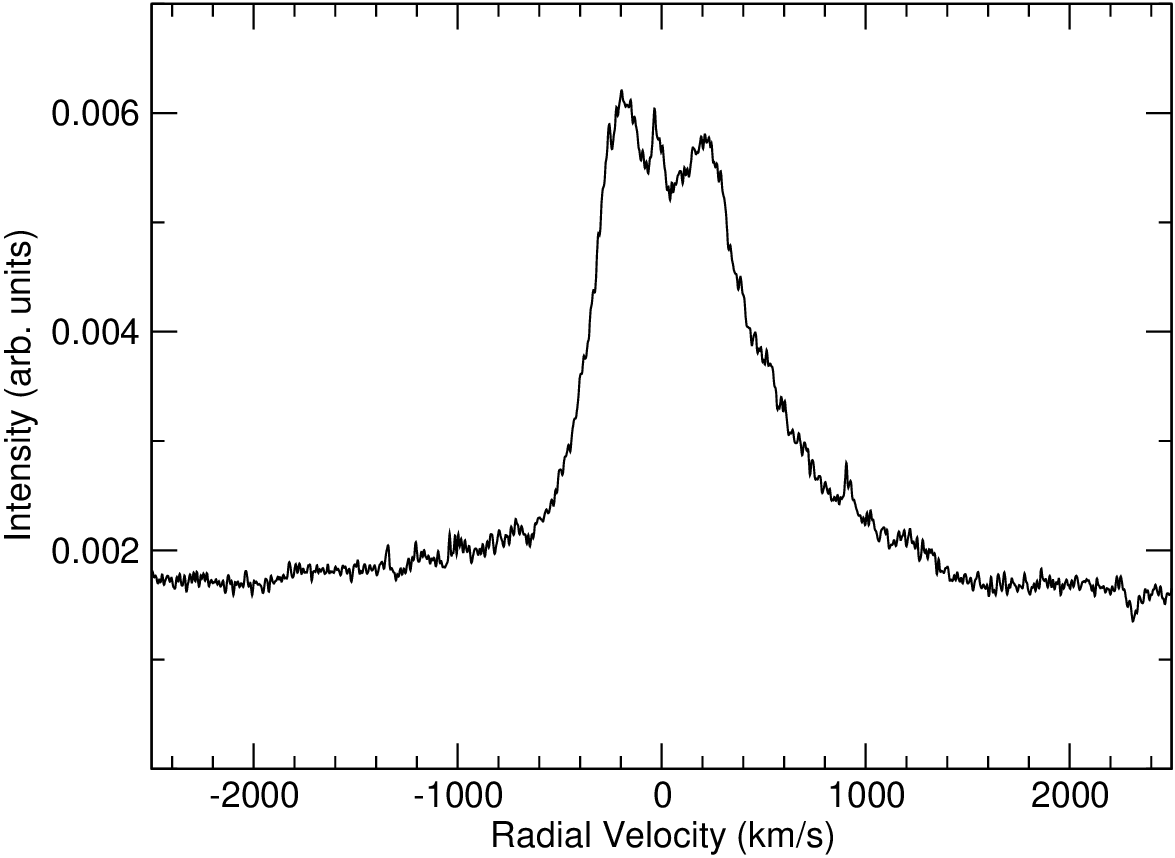}}
\caption{High resolution spectrum of H$\alpha$, showing its multiple peaks. \label{halfa}}
\end{figure}

As the high resolution spectrum was obtained with a 1.5 m telescope, it is quite noisy. Even so, we can identify a 
number of lines and obtain profile information (Table~\ref{lineprop}). The most intense hydrogen (H$\alpha$, 
H$\beta$) and helium (4686{\AA}) lines present a structure with three peaks. One, near rest wavelength, at about $+10$ km~s$^{-1}$,
and two side-peaks, with velocities of $-180$ km~s$^{-1}$ and $+240$ km~s$^{-1}$ (see Table~\ref{picos} and Fig.~\ref{halfa}).

\begin{figure}
\vspace*{30pt}
\centerline{\includegraphics[width=84mm]{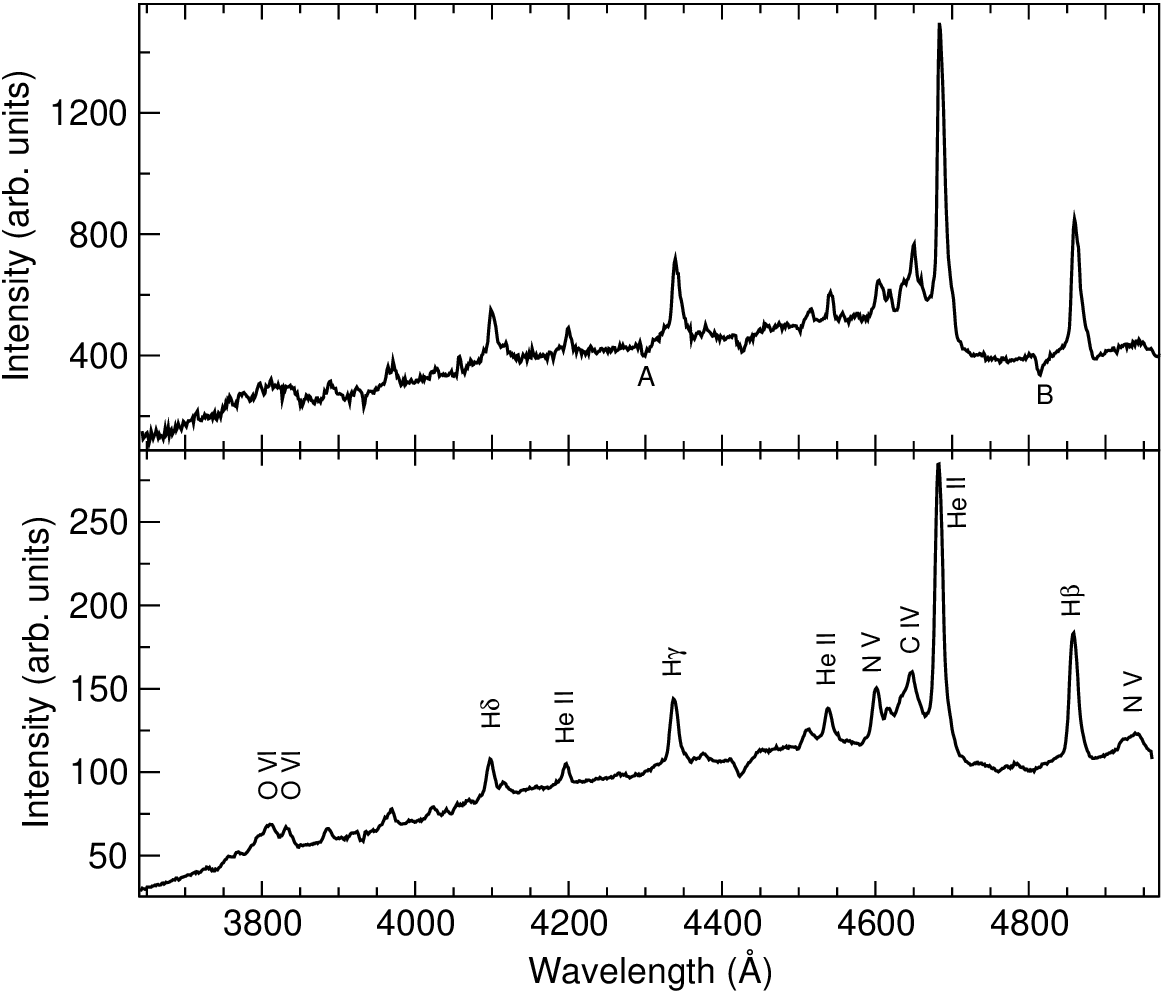}}
\caption{Average medium resolution spectrum of WX~Cen with the absorptions (labelled A and B, top spectrum) and without
	the absorptions (bottom spectrum). \label{jet}}
\end{figure}

Most of the lines can be identified with the transitions from the Balmer series, He II lines, C III, C IV,
N III, N V, O III, O V and O VI (see Table~\ref{lineprop}). But there are a few emission lines 
that are not generally seen in the context of Cataclysmic Variables or WR stars. We believe that they are spurious FEROS 
artifacts that appear in faint targets. There is no evidence of He I lines (see Table 4).

Another important feature in the high-resolution spectrum is the very high velocity associated to the extended red 
wing seen in the strongest Balmer and He II lines.

There are two additional points that deserve consideration. First, the high range of ionization of oxygen, which varies from 
O III to O VI (see Section~\ref{cnoab}). The second point is the absence of N IV emission, besides its analog of 
O V being clearly present.

\begin{table}
 \centering
% \begin{minipage}{140mm}
  \caption{Line identification in WX~Cen: High-resolution spectrum.  \label{lineprop}}
  \begin{tabular}{@{}llcc@{}}
  \hline
   Identif.     & $\lambda_{obs}$ ({\AA}) &  $-W_\lambda$ ({\AA})	& FWHM (km~s$^{-1}$)   \\
  \hline
  \multicolumn{4}{c}{Emission lines:} \\	\hline\\
He II/H$\delta$+N III	& 4101.8		& 7	& 700	\\	
He II/H$\gamma$		& 4341.3		& 9	& 900	\\
N III			& 4515			& ...                & ...	\\
He II			& 4539.6		& 12	& 1060 	\\
N V			& 4607.8		& 5	& 700	\\
N V			& 4625.9		& 3	& 840	\\
N III			& 4640			& 3           & 700	\\
C III			& 4650			& ...                 & ...	\\
C III+C IV		& 4664.1		& 3	& 450	\\
He II			& 4687.1		& 13	& 700	\\
He II/H$\beta$		& 4863.2		& 15	& 860	\\
O V		& 4930		& ...	& ...	\\
N V			& 4944		& ...                & ...	\\
O V		& 5115.7		& 2	& 300	\\
He II			& 5413.4		& 5	& 670	\\
O V		& 5580		& ...                 & ...	\\
O III			& 5592		& ...                 & ...	\\
O V		& 5600		& ...                 & ...	\\
C IV			& 5806.2		& 7           & 670	\\
C IV			& 5817.7		& 2           & 260	\\
He II/H$\alpha$		& 6563.8		& 44	& 820	\\ \hline
\multicolumn{4}{c}{Absorption features:} \\ \hline\\
Ca II			& 3933.49		& -0.76	& 66	\\
Ca II			& 3968.18		& -0.38	& 27	\\
DIB				& 5780.18		& -0.56	& 129	\\
Na I		& 5889.66		& -0.87	& 40	\\                                                                               
Na I 		& 5895.67		& -0.74	& 37	\\	        
\hline
\end{tabular}
%\end{minipage}
\end{table}

\begin{table}
 \centering
% \begin{minipage}{140mm}
  \caption{Velocities of the multiple peaks of emission lines, relative to the rest wavelengths. \label{picos}}
  \begin{tabular}{@{}lccc@{}}
  \hline
  	& peak 1 & peak 2 & peak 3   \\
  	&(km~s$^{-1}$)	&(km~s$^{-1}$)	&(km~s$^{-1}$)	\\
  \hline
He II 4686{\AA}	& $-161$	& +24	& +237	\\
H$\beta$	 		& $-186$	& $-13$	& +259	\\
He II 5411{\AA}	& $-176$	& +57	& +233	\\
C IV 5801{\AA}	& $-59:$	& +10:	& +269	\\
H$\alpha$		& $-185$	& $-21$	& +219	\\	
\hline
\end{tabular}
%\end{minipage}
\end{table}

\subsection{The medium-resolution spectrum}

\begin{figure}
\vspace*{50pt}
\centerline{\includegraphics[width=84mm]{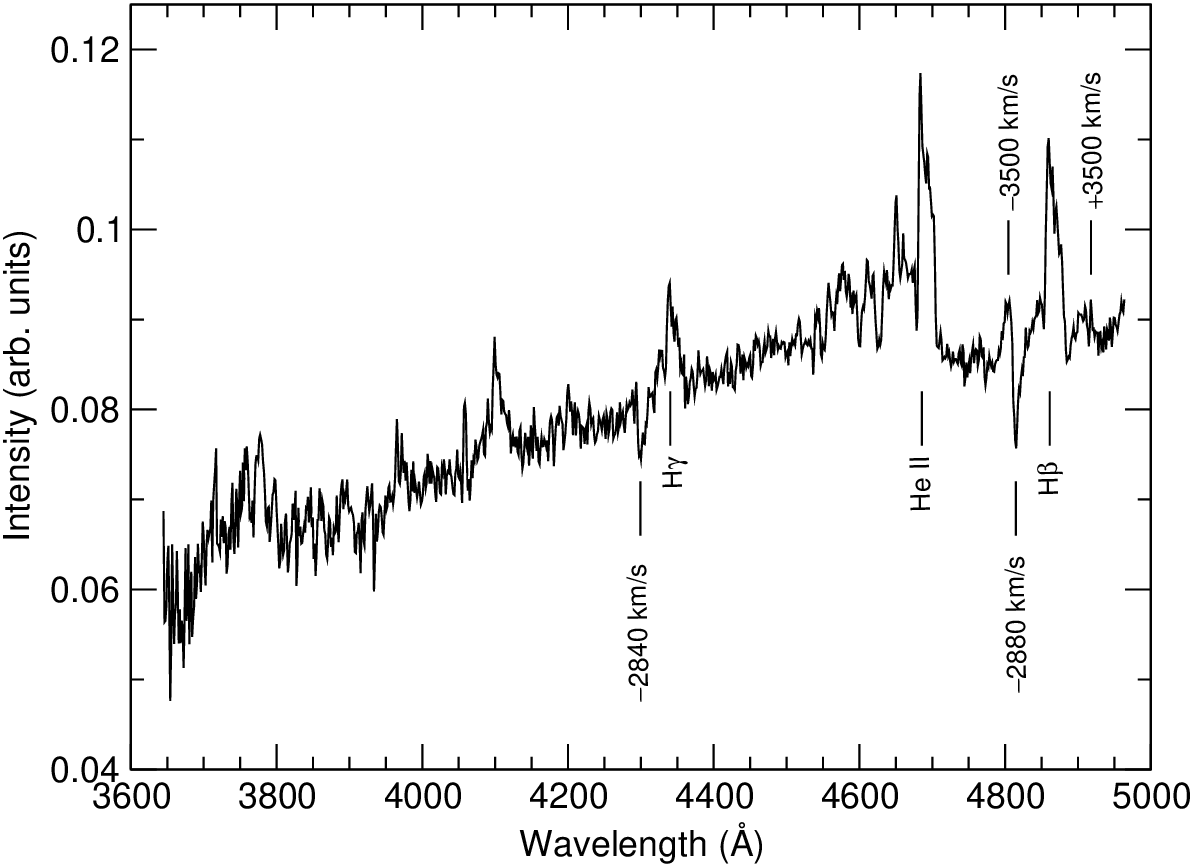}}
\caption{Average medium resolution spectrum of WX~Cen with absorption divided by the average spectrum 
	without absorption. \label{jet2} }
\end{figure}

In our medium-resolution spectra we see a unexpected phenomenon: on several occasions, absorptions are 
present in the Balmer lines blue wings, and this event is quite variable. Fig.~\ref{jet} illustrates the average spectra 
with and without the presence of such absorptions. In Fig.~\ref{jet2} we show the ratio of the average spectrum with 
absorption to the average spectrum without absorption. We clearly see the phenomenon. The absorption occurs 
at $-2880$ km~s$^{-1}$ in H$\beta$ and $-2840$ km~s$^{-1}$ in H$\gamma$. We see, at the same time, an emission in the bluest part of 
H$\beta$, at the velocity of $-3500$ km~s$^{-1}$. Evidence of a weak emission at $+3500$ km~s$^{-1}$ in H$\beta$ is also seen. 

\begin{table}
 \centering
 % \begin{minipage}{140mm}
  \caption{Line properties from medium-resolution spectrum.  \label{mediprop}}
  \begin{tabular}{@{}llcc@{}}
  \hline
   Identif.     & $\lambda_{obs}$ ({\AA}) &  $-W_\lambda$ ({\AA})	& FWHM (km~s$^{-1}$)   \\
  \hline
  \multicolumn{4}{c}{Emission lines:} \\	\hline\\
O III	& 3755-91	& ...	& ...	\\
O VI		& 3813.5		& ...	& ...	\\
O VI		& 3835.6		& ...	& ...	\\
He II/H$\epsilon$	& 3889.1		& 1.5	& 800	\\
He II		& 4025.5		& 0.9	& 790	\\
He II/H$\delta$	& 4101.2		& 4.8	& ...	\\
He II		& 4199.0		& 1.7 	& ...	\\
He II/H$\gamma$	& 4340.8		& 4.4	& 747	\\
N III		& 4515.3		& 1.1	& 963	\\
He II		& 4541.6		& 1.9	& 713	\\
N V		& 4604.0		& 3.1	& 730	\\
N V		& 4618.7		& 1.6	& 838	\\
C III-C IV/N III	& 4648.8	& 10.0	& 2000	\\
He II		& 4685.9		& 17.6	& 723	\\
He II/H$\beta$	& 4861.9		& 9.8	& 765	\\ \hline
\multicolumn{4}{c}{Absorption features:} \\ \hline\\
DIB 		& 4428.6		& -1.7	& ...	\\
Ca II	& 3933.4		& -0.4	& ...	\\

\hline
\end{tabular}
%\end{minipage}
\end{table}

Significant long-term variability in the line intensities and line ratios seems to be visible in spectra from 
distinct epochs. In table 1 from \citet{steiner98} one can see the equivalent widths of H$\alpha$,
H$\beta$ and He II 4686{\AA} as 61{\AA}, 15{\AA} and 37{\AA}. From our high-resolution spectra (Table~\ref{lineprop}),
we have, respectively, 37.0{\AA}, 13.8{\AA} and 12.2{\AA}.  Not only the lines seem to be much weaker but also the 
H$\beta$/He II ratio is much higher. This can easily be seen by comparing Fig.~\ref{jet} with fig.~1 in 
\citet{steiner98}. From our medium resolution spectra (Table~\ref{mediprop}), we can see that the intensity H$\beta$/He II 
ratio is again reversed with respect to our high resolution spectrum, although H$\beta$ is even weaker.

The lines also seem to be significantly narrower in the observations of the years 2000 and 2002 when compared 
to the observations of 1991 (DS95). While in 1991 the FWHM of the emission lines was typically 1100 km~s$^{-1}$, 
the latest observations show FWHM of about 700 km~s$^{-1}$. Even considering that the data were obtained with distinct 
instruments and different resolution, this difference seems to be significant. This suggests that in recent years the 
observations were made with the system in a less active state.

We have compiled all the lines observed in this object so far. These lines are shown in Table~\ref{cno},
 in which we list the
lines for similar terms from the relevant CNO ionization species. 

\begin{figure}
\vspace*{30pt}
\centerline{\includegraphics[width=84mm]{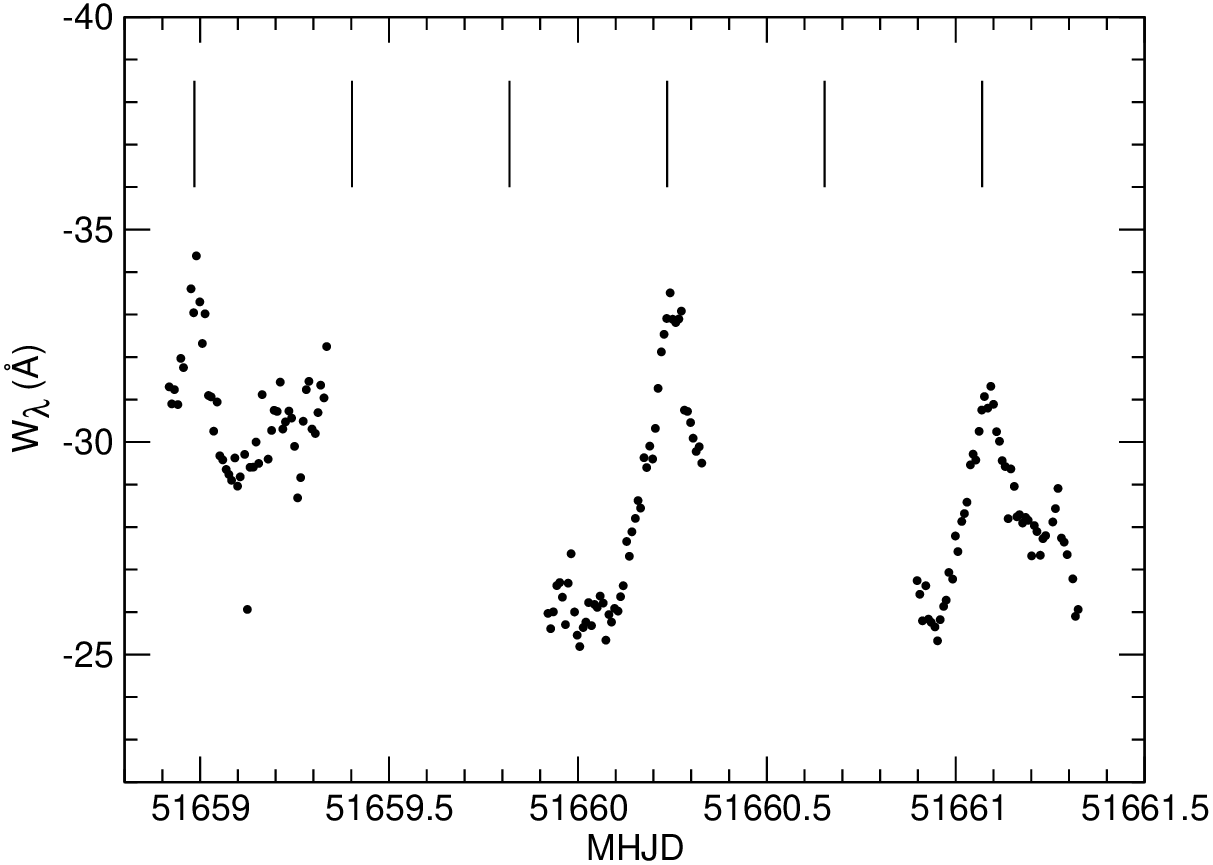}}
\caption{He II 4686{\AA} equivalent width measurements from the nights of 2000 March 24, 25 and 26.
	The vertical lines represent the timings of crossing from positive to negative values of the radial velocities 
	when compared to the system velocity, as defined in the spectroscopic ephemeris. \label{ew}  }
\end{figure}

Fig.~\ref{ew} shows the variation of the equivalent width as a function of time. It peaks at spectroscopic phase zero. 
This is quite similar to what has been reported for V617 Sgr \citep{cie}. The interpretation of 
this curve is quite straightforward: most of the eclipsed continuum comes from the heated surface of the 
secondary star (mass donor)  while the He II line is emitted in a non-eclipsed region (disc/wind/rim).

\subsection{The Temporal Variance Spectrum -- TVS}

\begin{figure}
\vspace*{20pt}
\centerline{\includegraphics[width=84mm]{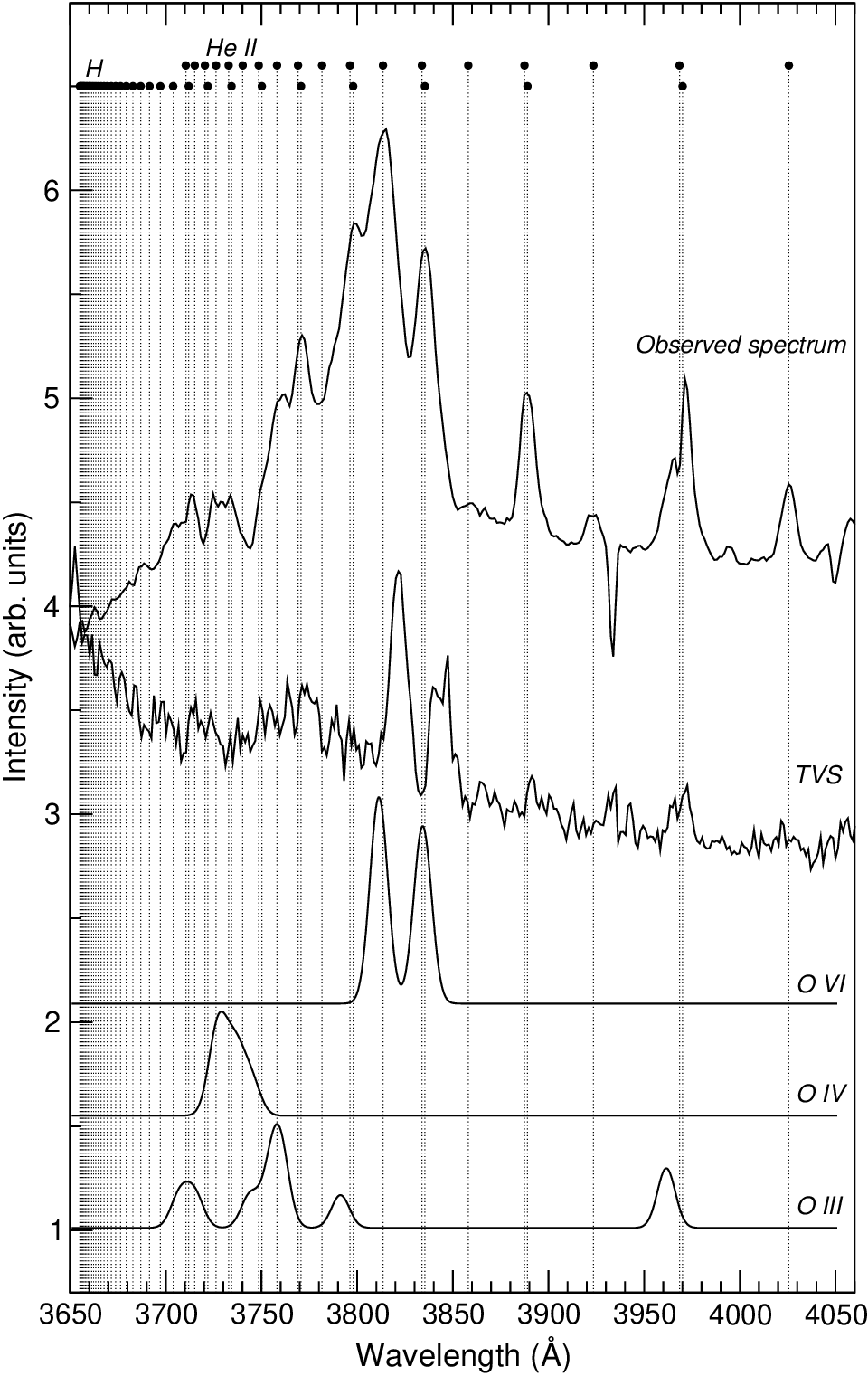}}
\caption{Intensity (observed) spectrum, TVS, and synthetic spectra for O VI, O IV and O III, from
	3650{\AA} to 4050{\AA}. Also shown (as vertical dotted lines) are the positions of the Pickering and Balmer 
	series lines. \label{tvs1} }
\end{figure}

\begin{figure}
\vspace*{15pt}
\centerline{\includegraphics[width=84mm]{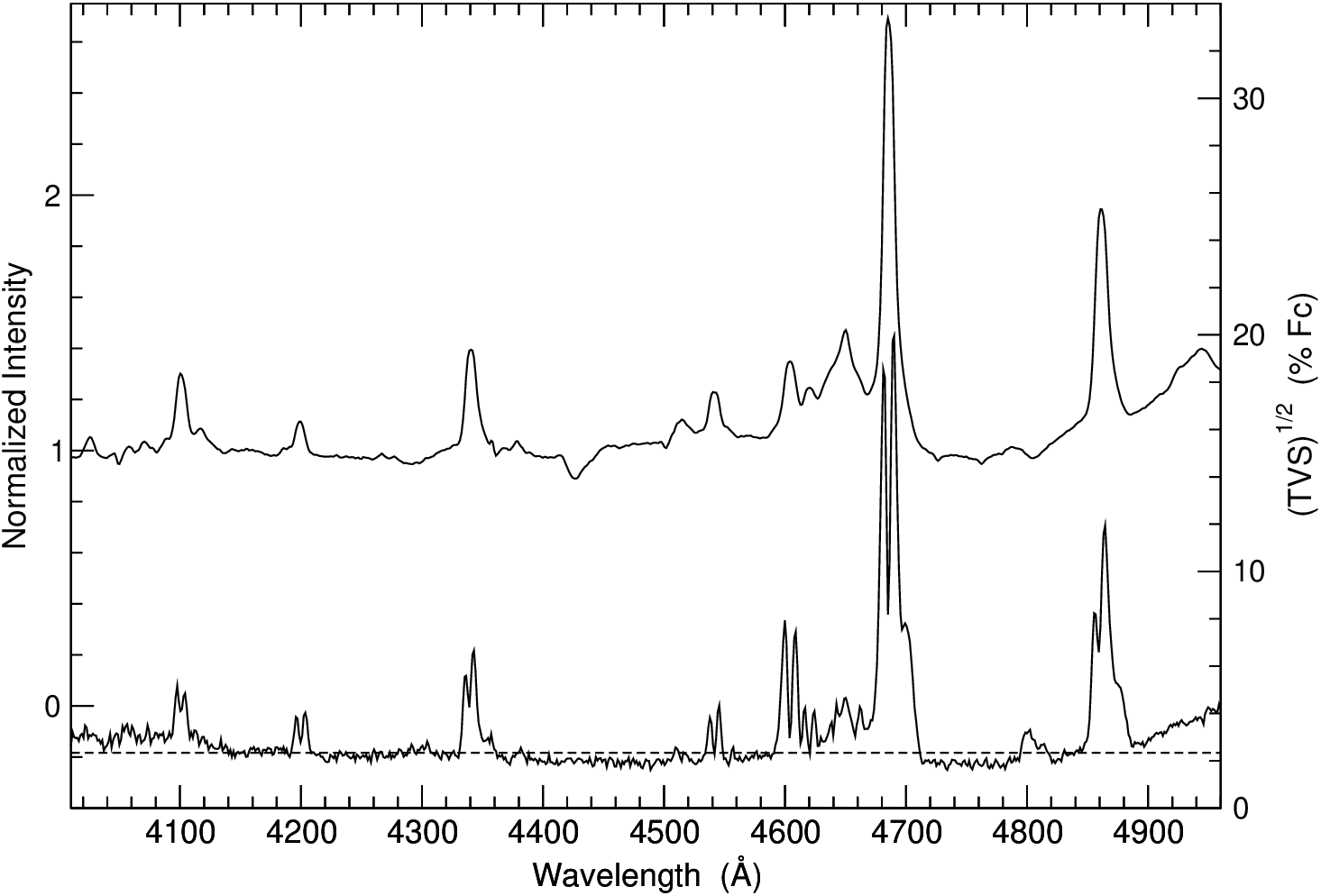}}
\caption{Intensity spectrum (above) and TVS (bellow) of WX~Cen from 4010{\AA} to 4950{\AA}. The TVS ordinate (right axis) gives the amplitudes
as percentage of the normalized continuum. The TVS statistical threshold for $p = 1\%$ is represented by the dashed line. \label{tvs2}  }
\end{figure}

In an attempt to examine the line variability in more detail we performed the Temporal Variance Spectrum
-- TVS -- analysis. In this procedure the temporal variance is calculated, for each wavelength pixel, from the
 residuals of the continuum normalized spectra to the mean spectrum. For further details and discussions of this method, see \citet{fulle}.
The observed median resolution (Cassegrain) mean spectrum of WX~Cen and the calculated TVS are shown in Figs.~\ref{tvs1}
and~\ref{tvs2}. A noticeable feature present in the N V 4603-19{\AA} and He II emission lines
is the double-peaked line profile, resulting not from a true line profile variation -- lpv -- but from a radial
velocity variation due to the binarity. This phenomenon is described in \citet{fulle} (see their fig.~1).
In this case we expect $\sigma/I \sim K/FWHM$, where $\sigma$ is the TVS line variance and $I$ is the related 
observed line intensity. The ratio $\sigma/I$ is 14 per cent for H$\beta$ and 12 per cent for He II 4686{\AA}.
These values of $\sigma/I$ are not an intrinsic property, but an artifact from the TVS analysis.

The TVS of WX~Cen also shows clear evidence of the variable absorptions in the blue
wings of the Balmer lines. There are two peaks, one with velocity of $-2900$ km~s$^{-1}$ and other with
$-3500$ km~s$^{-1}$, which correspond 
to the absorption and emission peaks near H$\beta$, respectively (see Fig.~\ref{jet2}).
Besides, the O VI 3811/34{\AA} lines show strong variability in 
the red wings, with a displacement of approximately 700 km~s$^{-1}$ relative to the rest wavelength. The red 
wings of the Balmer and 
He II lines show strong variability that extends to velocities up to $1500$ km~s$^{-1}$. This variability of the O VI lines, for instance, 
indicates that there are two emitting regions. One, with low variability, has normal velocity and another, highly variable, has 
high and positive velocity.  

\section{Discussion}
\subsection{The hydrogen content}

One way of analysing the presence/absence of hydrogen in a strong helium emitting spectrum is by comparing the 
He II 4859{\AA}+H$\beta$ to the geometric mean of the two adjacent transitions of the He II Pickering series. This criterion 
has been widely used, for instance, by  \citet{smit} in defining their three dimensional 
classification of WN stars. We define a Pickering parameter, $p$, such that

\eq
p = \frac{I(4859\mathrm{{\AA}} + 4861\mathrm{{\AA}})} {\left[I(4541\mathrm{{\AA}}) \times I(5411\mathrm{{\AA}}) \right] ^{1/2}}
\eeq

For a pure He II spectrum one expects $p=1$ (SSM96). Any value of $p$ larger than 1 would mean hydrogen in the object.
 In the case of WX~Cen, the measured value is about $p = 3.1$.
 
In case both hydrogen and helium lines were optically thin, it would be easy to determine the relative abundance 
between the two species. In this case  \citep{con3}:

 \eq
 \frac{N(H^{+})} {N(He^{++})} = p-1
 \eeq
 	
 \noindent and the abundance would be $N(H^{+})/N(He^{++})\sim2.1$.

The largest uncertainty in this determination comes from the hypothesis that all of the involved lines are optically thin.

For the optically thick case, one gets (see \citet{con3})

\eq
\frac{N(H^{+})} {N(He^{++})} = p^{3/2} -1
\eeq

\noindent which, for WX~Cen, gives
$N(H^{+})/N(He^{++})\sim4.5$. With a high resolution spectrum as is the present case, however,
we can determine the relative abundance using intensities at a given velocity in the profile, in which we can be 
reasonably secure that the lines are optically thin. For instance, if we use the intensity in the red wing that corresponds
to, say, 300 km~s$^{-1}$, we may be in such a situation. Defining a parameter $y(v)$, so that
 
 \eq
 \footnotesize y(v) = \frac{I4861\mathrm{{\AA}}(v) - [I4541\mathrm{{\AA}}(v+dv) \times I5411\mathrm{{\AA}}(v+dv)]^{1/2}} {[I4541\mathrm{{\AA}}
 (v) \times I5411{\mathrm{{\AA}} (v)]^{1/2} }}
 \eeq  
  		
\noindent where $I4861${\AA}$(v)$ is the intensity of the line at velocity $v$ from the rest wavelength of H$\beta$, $I5411${\AA}$ (v+dv)$ is the intensity 
of the line at velocity $v+dv$ from the He II 5411{\AA} rest wavelength. The $H^{+}/He^{++}$  abundance can be estimated as

\eq
\frac{N(H^{+})} {N(He^{++})} = y
\eeq

In the case of WX~Cen the parameter $y$ was measured at $v=300$ km~s$^{-1}$ while $dv=122$ km~s$^{-1}$. We found 
$N(H^{+})/N(He^{++}) \sim y \sim 2.3$. This is to be compared with the range found by DS95: $N(H^{+})/N(He^{++}) \sim $ 1.7 to 7.

In the \citet{smit} three dimensional classification scheme for WN stars, one could classify WX~Cen spectroscopically
as WN3hpec, where the pec suffix indicates, like in DI Cru, the presence of strong O VI emission lines. It is clear,
however, that we do not consider WX~Cen as having WR nature.

\subsection{CNO abundance} \label{cnoab}

An unusual characteristic we observe in the spectrum of WX~Cen is the large range of ionization degree. At the same 
time we see O VI emission, we also see O V, O IV and O III (Table~\ref{cno}).  This is very different from what is seen, for example, in WR stars. 
In these systems the wind and ionizing photons have similar dilution factors and the ionization parameter is constant to
a first approximation. Yet in V~Sagittae stars one expects a wind, accretion disc, hot spot, rim, spill-over and a secondary
star. The range of ionization parameter in each of these regions can be very distinct. The expected range of ionization 
species in V~Sge objects should, therefore, be larger than in WR stars.
\citet{her} also observed O III and O VI in V~Sge.

Is the N/O ratio anomalous in WX~Cen? Why do we see O V but not the same transition in N IV? From the intensity ratios,
we obtain N/O$<$0.1. This is similar to Planetary Nebulae of type II or to the solar abundance but is quite 
distinct from WN stars (N/O$\sim$ 20 to 50)

\subsection{The distance to the system}

\begin{figure}
%\vspace*{15pt}
\centerline{\includegraphics[width=84mm]{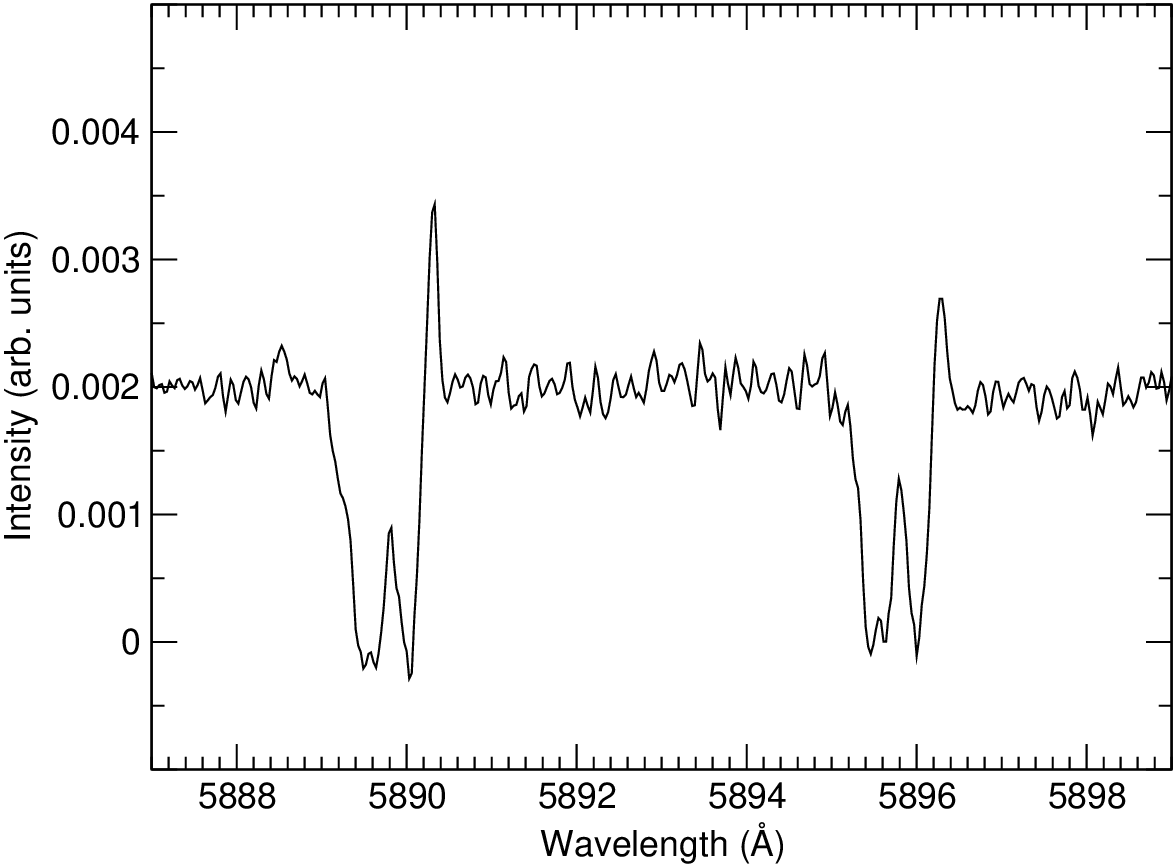}}
\caption{Na I interstellar absorption lines. The red wing emissions are due to bad sky subtraction.
	   \label{sodio}}
\end{figure}

DS95 made a distance determination on the basis of the diffuse interstellar band --DIB--  at 4430{\AA}, interstellar Na D lines
and the decrement of He II lines and found a distance of about 1.4 kpc.  To measure the equivalent width of the 4430{\AA} 
band is difficult, given that it is hard to define the nearby continuum, and the Na D lines are saturated (Fig.~\ref{sodio}).
We will, therefore, revisit this issue. Let's first estimate the interstellar extinction from the diffuse interstellar bands. The DIBs and 
reddening have been calibrated by many authors. From \citet{herb2}, we have

\eq
E(B-V) = W_{\lambda}(4430\mathrm{\AA})/2.3 ~~~ \mathrm{mag}
\eeq

This is different from the relation in \citet{allen}, used by DS95, by about a factor of 2. We found  
$W_{\lambda}(4430\mathrm{\AA}) = 1.7$ (Table~\ref{mediprop}), and obtain $E(B-V) = 0.74$.

The most reliable DIB for evaluating the interstellar extinction in the case of 
WX Cen seems to be the 5780{\AA} feature. We measured an equivalent width of 
$W_{\lambda} =0.56 \pm 0.01${\AA}. \citet{somer} gives the following relation between its  equivalent 
width and color excess
 
\eq
E(B-V) = (W_{\lambda}(5780\mathrm{\AA})  + 46) / 940
\eeq

\noindent where $W_{\lambda}$ is given in m{\AA}. In the case of WX Cen we have $E(B-V) = 0.63$ and we will 
adopt this as the color excess. This is about 50 per cent larger than the value estimated 
by DS95. From a calibration of the color excess \textit{versus} distance, obtained both for 
open clusters \citep{wilt} and field stars (see Table~\ref{aglom}), we derive a distance of about $2.8 \pm 0.3$ kpc.

The Na I lines for WX Cen and stars in its field \citep{fran} have velocities, in LSR, that are listed in Table~\ref{navel}.
Four absorption systems can be identified and we will call these components from 1 to 4.

\begin{table*}
 \centering
\begin{minipage}{140mm}
  \caption{Radial velocities (km~s$^{-1}$) of Na ID absorption components and distances (kpc).  \label{navel}}
  \begin{tabular}{@{}cccccc@{}}
  \hline
  Component & 	WX~Cen & SAO252025 & SAO252085 & SAO252146 & Coalsack \\
  \hline
1	& -4.1	& -0.8	& -3.5	& -4.8	& -2/-4	\\
2	& -23.9	& -26.6	& -21.6	& -23.8     & ...	\\
3	& -32.0	& -37	& -33.1	& -35.1     & ...	\\
4               & -39.0	& 	& -42	&  ...	&    ...	\\                  \hline\\
distance	&  ...	& 1.81	& 2.48	& 2.72	& 0.2	\\
\hline
\end{tabular}
\end{minipage}
\end{table*}

From the other stars in the field of the Coalsack, observed by \citet{fran}, one 
concludes that the distance to WX Cen has to be well in excess of 800 pc. The numbers 
in Table~\ref{navel}, as well as the equivalent widths of the absorption components 2 and 3, 
suggest a distance of about 1.8 to 2.7 kpc which seems to be consistent with the value 
derived above.

The absolute magnitude is, then, of about $M_V = - 0.5$ and color $(B-V)_0 = -0.2$.

\subsection{Spill-over; SN~Ia precursor?}

A somewhat striking characteristic of the spectrum of WX Cen are the red wings in the emission lines, 
seen in the high resolution spectrum, extending to about $+1300$ km~s$^{-1}$. This is similar to what was 
observed by \citet{cie} in V617 Sgr. In the Doppler diagram presented by these authors, 
after subtracting the symmetric profile, one can notice a high velocity component extending to 
$V \sim -800$ to  $-1000$  km~s$^{-1}$, both in $V_x$ and in $V_y$. In addition there is a strong emission concentrated at 
$V_x \sim -200$ km~s$^{-1}$. \citet{steiner99} explained these characteristics by assuming that the $-200$ km~s$^{-1}$ 
component is due to the high disc rim, similar to the ones observed in CBSSs \citep{meyer}. 
The high velocity component was explained in V617 Sgr as a consequence of spill-over gas, following 
ballistic trajectories and closely approaching the white dwarf. Such a configuration would explain the red 
wing characteristic if the high resolution spectrum was taken at a phase immediately before the 
photometric minimum. This is indeed the case. The central exposure of the high resolution spectrum 
was at photometric phase 0.787. We predict that at opposite phase the wing should be to the blue. This 
phenomenon is also likely to be variable as there is no hint of such a component in the 
Doppler diagram presented in DS95.

A puzzling feature still remains unexplained. The TVS spectrum shows a highly 
variable component in the highly ionized species O VI. This component is seen at $V \sim 700$ km~s$^{-1}$. 
This component is also seen in the TVS red wings of H$\beta$ and He II 4686{\AA}. We have no 
explanation for this. 

The satellite-like events at the Balmer lines, being interpreted as due to ejected gas, are consistent 
with the idea that the compact star is a white dwarf. With an orbital inclination of $i=55\degr$, and 
observed velocity of 3500 km~s$^{-1}$, we derive the intrinsic velocity of the ''jet'' as $V=6100$ km~s$^{-1}$. 
This corresponds to the escape velocity of a white dwarf with a mass of $M \sim 0.9 M_{\sun}$. With the 
high accretion rate ($dM_1/dt \sim 10^{-7} M_{\sun}~year^{-1}$)
required in order to have surface nuclear burning \citep{kah}, this white dwarf may reach the Chandrasekhar
 limit and become a  SN~Ia on a time-scale of $5 \times 10^{6}$ years.

CBSSs have been considered as possible candidates to SN~Ia precursors \citep{hach}. These 
authors have shown a channel to SN~Ia by evolving a helium rich Supersoft X-ray binary. 
\citet{kato} also showed that V Sge may become a SN~Ia.

\section{Conclusions} 

	The main conclusions of this paper are:
 
\begin{enumerate}

\item We confirmed the orbital period of WX~Cen determined by DS95 and refined its value to $P_{orb} = 0.416~961~5 (\pm 22)$ d, based
on spectroscopic observations. 

\item The light curve has total amplitude of approximately 0.32 magnitudes and is non-sinusoidal in the sense 
of having a narrow, V-shaped, minimum. The object presents flickering with time-scales of tens of minutes. 
Night-to-night variations of about 0.08 mag are also observed.

\item We identified most of the emission lines as due to Balmer, He II, C IV, N V, O III, O V, O VI.

\item The object shows absorption satellites in the Balmer lines with $V = -2900$ km~s$^{-1}$ and in emission with 
$V = \pm 3500$ km~s$^{-1}$. These highly variable events remind the satellites in emission in CBSS. 

\item The analysis of the emission lines show that the He II 4686{\AA} line became weaker between 1991 and 
2000/2002 observations. At the same time the emission lines were narrower, suggesting that in 2000/03 the system 
was in a less active state than in 1991.

\item The object presents red-shifted emission in high ionization species.  Extended red wing emission is 
also seen in the strongest lines. The O VI lines show strong variability with velocity of about 700 km~s$^{-1}$.

\item An analysis of the He II Pickering series decrement shows that the system has significant amount of 
hydrogen, with an abundance, by number, between 2.1 and 4.5.

\item We estimate the color excess as $E(B-V)=0.63$  on the basis of the observed DIB at 5780{\AA}. 
Given the distance-color excess relation in the direction of WX Cen, this implies a distance of $2.8 \pm 0.3$ kpc. 
Interstellar absorption of the Na I D lines show components at $-4.1$ km~s$^{-1}$, which corresponds to the velocity 
of the Coalsack, and three other components a $-23.9$, $-32.0$ and $-39.0$ km~s$^{-1}$. These components are also seen 
with similar strengths in field stars that have distances between 1.8 and 2.7 kpc. The intrinsic $B-V$ color index 
of WX Cen is $(B-V)_0=-0.2$ and the absolute magnitude, $M_V = -0.5$.

\item A possible explanation for the extended wings in the strong emission lines is that  
the system has a spill-over stream similar to what is seen in V617~Sgr. We predict that when observed in opposite 
phase, blue wings would be observed.

\item The velocity of the satellites is consistent with the idea that the central star is a white dwarf with a mass 
of $M \sim 0.9 M_{\sun}$. With the high 
accretion rate, it may become a  SN~Ia in a time-scale of $5 \times 10^{6}$ years.

\end{enumerate}

\appendix

\section{}

\clearpage

\begin{table*}
 \centering
 \begin{minipage}{140mm}
  \caption{The CNO emission line table for WX~Cen.  \label{cno}}
  \begin{tabular}{@{}lccccc@{}}
\hline
   & \bf{O}	&	& \bf{N}	&	& \bf{C}	\\
\hline
  	& \bf{O III}	&	&	&	& 	\\
IP (eV)	& 54.94		&	&	&	&	\\
[5pt]	
\textbf{Term}	& {\boldmath $\lambda$}\textbf{({\AA})} -- {\boldmath $W_{\lambda}$} \textbf{({\AA})}&	&	&	&	\\
$3s~^{1}P^{0}$~--~$3p~^{1}P$ & 5592  \hspace{\stretch{1}} y\footnote{Meaning of the codes: y = line present; 
bl = blended line; np = line not present.}	&	&	&	& \\
[2pt]
$3s~^{3}P^{0}$~--~$3p~^{3}D$ & 3760  \hspace{\stretch{1}}  y	&	&	&	& \\
[2pt]
			    & 3755  \hspace{\stretch{1}}  y	&	&	&	& \\
[2pt]
			    & 3757  \hspace{\stretch{1}}  y	&	&	&	& \\
[2pt]
			    & 3791  \hspace{\stretch{1}}  y?	&	&	&	& \\
[10pt]	  
 	& \bf{O IV}	&	& \bf{N III}	&	&	\\
IP (eV)	& 77.41		&	& 47.45		&	&	\\ 
[5pt]
\textbf{Term}	& {\boldmath $\lambda$}\textbf{({\AA})} -- {\boldmath $W_{\lambda}$}\textbf{({\AA})}	&	&	
{\boldmath $\lambda$}\textbf{({\AA})} -- {\boldmath $W_{\lambda}$} \textbf{({\AA})} &	&	\\
Doublets	&	&	&	&	&	\\
$3s~^{2}S$~--~$3p~^{2}P^{0}$  & 3063  \hspace{\stretch{1}} ...	&	& 4097\hspace{\stretch{1}}bl &	&	\\
[2pt]
			     & 3072  \hspace{\stretch{1}}...	&	& 4103\hspace{\stretch{1}}...	&	&	\\
[2pt]
$3s~^{2}P^{0}$~--~$3p~^{2}D$  & 3349  \hspace{\stretch{1}}...	&	& 4200\hspace{\stretch{1}}bl &	&	\\
[2pt]
			     & 3348  \hspace{\stretch{1}}...	&	& 4196\hspace{\stretch{1}} ...   &	&	\\
[2pt]
			     & 3378  \hspace{\stretch{1}}...	&	& 4216\hspace{\stretch{1}}  ...  &	&	\\
[2pt]
$3p~^{2}P^{0}$~--~$3d~^{2}D$  & 3412  \hspace{\stretch{1}}...	&	& 4641\hspace{\stretch{1}}  y  &	&	\\
[2pt]
			     & 3404  \hspace{\stretch{1}}...	&	& 4634\hspace{\stretch{1}}  y  &	&	\\
[2pt]
			     & 3414  \hspace{\stretch{1}}...	&	& 4642\hspace{\stretch{1}}  y?  &	&	\\
[2pt]
Quadruplets	&	&	&	&	&	\\
$3s~^{4}P^{0}$~--~$3p~^{4}D$ & 3386  \hspace{\stretch{1}}...	&	& 4515\hspace{\stretch{1}}1.1	&	&	\\
[2pt]	
			    & 3381  \hspace{\stretch{1}}...	&	& 4511\hspace{\stretch{1}}...		&	&	\\
[2pt]
$4p~^{2}D$~--~$4d~^{2}P^{0}$ & 7716  \hspace{\stretch{1}}...	&	&	... \hspace{\stretch{1}}			&	&	\\
[2pt]
$5f~^{2}F^{0}$~--~$6g~^{2}G$ &	...	\hspace{\stretch{1}}...		&	 &  8019\hspace{\stretch{1}}...	&	&	\\
[10pt]
	& \bf{O V}	&	& \bf{N IV}	&	& \bf{C III}	\\
IP (eV)	& 113.90		&	& 77.47		&	& 47.89		\\ 
[5pt]
\textbf{Term}	& {\boldmath $\lambda$}\textbf{({\AA})} -- {\boldmath $W_{\lambda}$}\textbf{({\AA})}	&	&
{\boldmath $\lambda$}\textbf{({\AA})} -- {\boldmath $W_{\lambda}$} \textbf{({\AA})} &	&
{\boldmath $\lambda$}\textbf{({\AA})} -- {\boldmath $W_{\lambda}$}\textbf{({\AA})}	 \\
Singlets	&	&	&	&	&	\\
$3s~^{1}S$~--~$3p~^{1}P^{0}$	& 5114  \hspace{\stretch{1}} y	&	& 6381\hspace{\stretch{1}}...		&	&
8500\hspace{\stretch{1}} ...	\\
[2pt]
Triplets	&	&	&	&	&	\\
$3s~^{3}S$~--~$3p~^{3}P^{0}$	& 2781\hspace{\stretch{1}}...		&	& 3479\hspace{\stretch{1}}...		&	& 4647\hspace{\stretch{1}}10: \\
[2pt]
			  	& 2787\hspace{\stretch{1}}...		&	& 3482\hspace{\stretch{1}}...&	& 4650\hspace{\stretch{1}}...	\\
[2pt]
			  	& 2790\hspace{\stretch{1}}...	&	& 3484\hspace{\stretch{1}}...	&	& 4651\hspace{\stretch{1}}...	\\
[2pt]
$3s~^{3}P^{0}$~--~$3p~^{3}P$	& 2731\hspace{\stretch{1}}...	&	& 3463\hspace{\stretch{1}}...	&	& 4666\hspace{\stretch{1}}2.3: \\
[2pt]
				& 2744\hspace{\stretch{1}}...	&	& 3461\hspace{\stretch{1}}...	&	& 4673\hspace{\stretch{1}} ...\\
[2pt]
				& 2729\hspace{\stretch{1}}...	&	& 3475\hspace{\stretch{1}}...	&	& 4663\hspace{\stretch{1}} ...\\
[2pt]
$3p~^{3}P^{0}$~--~$3d~^{3}D$	& 5598\hspace{\stretch{1}}y&	& 7123\hspace{\stretch{1}}np &	& 9715\hspace{\stretch{1}}... \\
[2pt]
				& 5580\hspace{\stretch{1}}y&	& 7109\hspace{\stretch{1}}np &	& 9705\hspace{\stretch{1}}...	\\
[2pt]
				& 5573\hspace{\stretch{1}}y&	& 7103\hspace{\stretch{1}}np &	& 9701\hspace{\stretch{1}}...	\\
[2pt]
				& 5604\hspace{\stretch{1}}...	&	& 7127\hspace{\stretch{1}}...	&	& 9718\hspace{\stretch{1}}...	\\
[10pt]
\hline
\end{tabular}
\end{minipage}
\end{table*}

\begin{table*}
 \centering
 \begin{minipage}{140mm}
  \contcaption{}
    \begin{tabular}{@{}lccccc@{}}
\hline
   & \bf{O}	&	& \bf{N}	&	& \bf{C}	\\
\hline
		& \bf{O VI}	&	& \bf{N V}	&	& \bf{C IV}	\\
IP (eV)		& 138.12		&	& 97.89		&	& 64.49		\\ 
[5pt]
\textbf{Term}	& {\boldmath $\lambda$}\textbf{({\AA})} -- {\boldmath $W_{\lambda}$}\textbf{({\AA})}	&	&
{\boldmath $\lambda$}\textbf{({\AA})} -- {\boldmath $W_{\lambda}$}\textbf{({\AA})}	&	&
{\boldmath $\lambda$}\textbf{({\AA})} -- {\boldmath $W_{\lambda}$} \textbf{({\AA})}\\
$3s~^{2}S$~--~$3p~^{2}P^{0}$	& 3811\hspace{\stretch{1}}y &	& 4604\hspace{\stretch{1}}3.4 &	& 5801\hspace{\stretch{1}}3.9 \\
[2pt]
				& 3834\hspace{\stretch{1}}y &	& 4620\hspace{\stretch{1}}2.8 &	& 5812\hspace{\stretch{1}}2.4 \\
[2pt]
$4s~^{2}S$~--~$4p~^{2}P^{0}$ 	& 9342\hspace{\stretch{1}}	&	& 11331\hspace{\stretch{1}}... &	& 14335\hspace{\stretch{1}} ...\\
[2pt]
				& 9398\hspace{\stretch{1}}...	&	& 11374\hspace{\stretch{1}} ...&	& 14362\hspace{\stretch{1}} ...\\
[2pt]
(5-6)	& 2070	\hspace{\stretch{1}}...	&	& 2981	  \hspace{\stretch{1}}	...&	& 4658	\hspace{\stretch{1}}y?	\\
[2pt]
(6-7)	& 3435	\hspace{\stretch{1}}...	&	& 4945	  \hspace{\stretch{1}}y	&	& 7726	\hspace{\stretch{1}}y	\\
[2pt]
(6-8)	& 2083	\hspace{\stretch{1}}...	&	& 2998	  \hspace{\stretch{1}}...	&	& 4685	\hspace{\stretch{1}}bl 	\\
[2pt]
(7-8)	& 5291	\hspace{\stretch{1}}y?	&	& 7618	  \hspace{\stretch{1}}y	&	& 11908	\hspace{\stretch{1}}...	\\
[2pt]
(7-9)	& 3143	\hspace{\stretch{1}}...	&	& 4520	  \hspace{\stretch{1}}...	&	& 7063	\hspace{\stretch{1}}y?	\\
[2pt]
(8-9)	& 7715	\hspace{\stretch{1}}y	&	& 11110	  \hspace{\stretch{1}}...	&	& 17368	\hspace{\stretch{1}}...	\\
[2pt]
(7-10)	& 2431	\hspace{\stretch{1}}...	&	& 3502	  \hspace{\stretch{1}}	...&	& 5471	\hspace{\stretch{1}}np	\\
[2pt]
(8-10)	& 4494	\hspace{\stretch{1}}...	&	& 6478	  \hspace{\stretch{1}}...	&	& 10124	\hspace{\stretch{1}}...	\\
[2pt]
(9-10)	& 11033	\hspace{\stretch{1}}...	&	& 15536	  \hspace{\stretch{1}}	...&	& 24278	\hspace{\stretch{1}}...	\\
[2pt]
(8-11)	& 3427	\hspace{\stretch{1}}...	&	& 4943	  \hspace{\stretch{1}}...	&	& 7737	\hspace{\stretch{1}}y?	\\
[2pt]
(9-11)	& 6202	\hspace{\stretch{1}}...	&	& 8927	  \hspace{\stretch{1}}...	&	& 13954	\hspace{\stretch{1}}	...\\
[2pt]
(10-11)	& 14590	\hspace{\stretch{1}}...	&	& 21000	  \hspace{\stretch{1}}...	&	& 32808	\hspace{\stretch{1}}...\\ 
\hline
\end{tabular} 
\end{minipage}
\end{table*}

%\tablecomments{Meaning of the codes: y = line present; bl = blended line; wk = weak line; np = line not present; tell = telluric line.}                      
%\tablerefs{\footnotesize IR: \citet{schmu}; \citet{vre}. Optical: Vee02a; CSH95; \citet{vre}. UV: CSH95; \citet{vacca}. 
%Rest wavelengths and line identifications were obtained from: NIST Atomic Spectra Database (http://www.physics.nist.gov/cgi-bin/AtData/main$_{-}$asd),
%Atomic Line List (http://www.pa.uky.edu/$\sim$peter/atomic/) 
%and Atomic Molecular and Optical Database Systems (http://amods.kaeri.re.kr/).	

\begin{table*}
 \centering
 \begin{minipage}{140mm}
%  \contcaption{}
 \caption{Distance and color excess for open clusters and stars in the fields of WX~Cen.  \label{aglom}}
  \begin{tabular}{@{}lccccl@{}}
\hline
 Name  &		\textit{V}&	\textit{B-V}&	\textit{E(B-V)}&	d (kpc)&	Sp\\
\hline
\multicolumn{6}{c}{Clusters} \\ \hline \\
Stock 16          &...    &...    &0.49   &1.64&...	  \\
Ruprecht 107	  &...    &...    &0.46   &1.44&...	  \\
Collinder 271	  &...    &...    &0.29   &1.17&...	  \\
Basel 18	  &...    &...    &0.51   &2.23&...	  \\
NGC 4815	  &...    &...    &0.81   &3.08&...	  \\
Trumpler 21	  &...    &...    &0.20   &1.26&...	  \\
BH 144		  &...    &...    &0.60   &9.35&...	  \\
Hogg 16		  &...    &...    &0.41   &1.58&...	  \\
Collinder 272	  &...    &...    &0.47   &2.04&...	  \\
Pismis 18	  &...    &...    &0.50   &2.24&...	  \\  \hline 
\multicolumn{6}{c}{Stars} \\ \hline\\
114800		   &8.00   &0.12   &0.35   &1.17   &	   B2 IV NE\\
114737		   &8.01   &0.17   &0.46   &2.00   &	   O9 IV   \\
114886 		   &6.87   &0.12   &0.40   &1.23   &	   O9 IV      \\
114478		   &8.68   &0.49   &0.65   &3.21   &	   B1 Ib       \\
115034		   &8.80   &0.09   &0.35   &1.90   &	   B0.5 Vn  \\
114122		   &8.57   &0.59   &0.73   &3.10   &	   B1 Iab    \\
115363		   &7.79   &0.61   &0.74   &3.20   &	   B1 Ia       \\
113754		   &9.50   &0.63   &0.90   &4.43   &	   O9.5 Ia   \\
CP-623111	   &10.20  &0.19   &0.44   &2.69   &	   B1 V       \\
CP-613512 	   &10.03  &0.35   &0.58   &3.34   &	   B1 III	\\
115071		   &7.96   &0.24   &0.50   &1.03   &	   B0.5 Vn\\
113432		   &8.86   &0.80   &0.91   &2.35   &	   B1 1b    \\
CP-632544 	   &9.97   &0.49   &0.67   &5.64   &	   B1 Ib     \\
CP-613598 	   &10.3   &0.84   &0.96   &6.40   &	   B5 Iap   \\
113411		   &9.05   &0.46   &0.71   &2.72   &	   O9.5 IV \\
WX Cen		   &13.6/14.2		   &0.4    &0.63   & $2.9\pm 0.3$ &	   WN3h pec    \\	       

\hline
\end{tabular}
\end{minipage}
\end{table*}

\label{lastpage}

\end{document}